\documentclass[twocolumn,amsmath,amssymb,aps]{revtex4}

\usepackage{epsfig}
\usepackage{graphicx}
\usepackage{dcolumn}
\usepackage{bm}
\usepackage{wasysym}
\usepackage{marvosym}
\usepackage{times,amsmath,amssymb}
\usepackage{color}


\begin{document}

\title{Making graphene visible}

\author{P. Blake, E. W. Hill}
\affiliation{Department of Computer Sciences, University of Manchester, Manchester, M13~9PL, UK}

\author{A. H. Castro Neto}

\affiliation{Department of Physics, Boston University, 590 Commonwealth Avenue, Boston, MA 02215,USA}

\author{K. S. Novoselov, D. Jiang, R. Yang, T. J. Booth, A. K. Geim}
\affiliation{Department of Physics and Astronomy, University of Manchester, Manchester, M13~9PL, UK}


\begin{abstract}
Microfabrication of graphene devices used in many experimental studies currently relies on the fact that graphene crystallites can be visualized using optical microscopy if prepared on top of Si wafers with a certain thickness of SiO$_2$. We study graphene's visibility and show that it depends strongly on both thickness of SiO$_2$ and light wavelength. We have found that by using monochromatic illumination, graphene can be isolated for any SiO$_2$ thickness, albeit 300~nm (the current standard) and, especially, $\approx$100~nm are most suitable for its visual detection. By using a Fresnel-law-based model, we quantitatively describe the experimental data.
\end{abstract}

\maketitle

Since it was reported in 2004~\cite{kostya1}, graphene---a one-atom-thick flat allotrope of carbon---has been attracting increasing interest~\cite{kostya1,nmat,neto_pw}. This interest is supported by both the realistic promise of applications and the remarkable electronic properties of this material. It exhibits high crystal quality, ballistic transport on a submicron scale (even under ambient conditions) and its charge carriers accurately mimic massless Dirac fermions~\cite{nmat,neto_pw,kim}. Graphene samples currently used in experiments are usually fabricated by micromechanical cleavage of graphite: a euphemism for slicing this strongly layered material by gently rubbing it against another surface~\cite{pnas}. The ability to create graphene with such a simple procedure ensures that graphene was produced an uncountable number of times since graphite was first mined and the pencil invented in 1565~\cite{pencil}.

Although graphene is probably produced every time one uses a pencil, it is extremely difficult to find small graphene crystallites in the `haystack' of millions of thicker graphitic flakes which appear during the cleavage. In fact, no modern visualization technique (including atomic-force, scanning-tunneling and electron microscopies) is capable of finding graphene because of their extremely low throughput at the required atomic resolution or the absence of clear signatures distinguishing atomic monolayers from thicker flakes. Even Raman microscopy, which recently proved itself as a powerful tool for distinguishing graphene monolayers,~\cite{ferrari} has not yet been automated to allow search for graphene crystallites. Until now, the only way to isolate graphene is to cleave graphite on top of an oxidized Si wafer and then carefully scan its surface in an optical microscope. Thin flakes are sufficiently transparent to add to an optical path, which changes their interference color with respect to an empty wafer~\cite{kostya1}. For a certain thickness of SiO$_2$, even a single layer was found to give sufficient, albeit feeble, contrast to allow the huge image-processing power of the human brain to spot a few micron-sized graphene crystallites among copious thicker flakes scattered over a mm-sized area.

So far, this detection technique has been demonstrated and widely used only for a SiO$_2$ thickness of 300~nm (purple-to-violet in color) but a 5${\%}$ change in the thickness (to 315nm) can significantly lower the contrast~\cite{nmat}. Moreover, under nominally the same observation conditions, graphene's visibility strongly varies from one laboratory to another (e.g. see images of single-layer graphene in Refs~\cite{kostya1,kim}), and anecdotal evidence attributes such dramatic differences to different cameras, with the cheapest ones providing better imaging~\cite{monochrome-cheapest}. Understanding the origin of this contrast is essential for optimizing the detection technique and extending it to different substrates, aiding experimental progress in the research area.

\begin{figure}[htb!]
\begin{center}\includegraphics[]{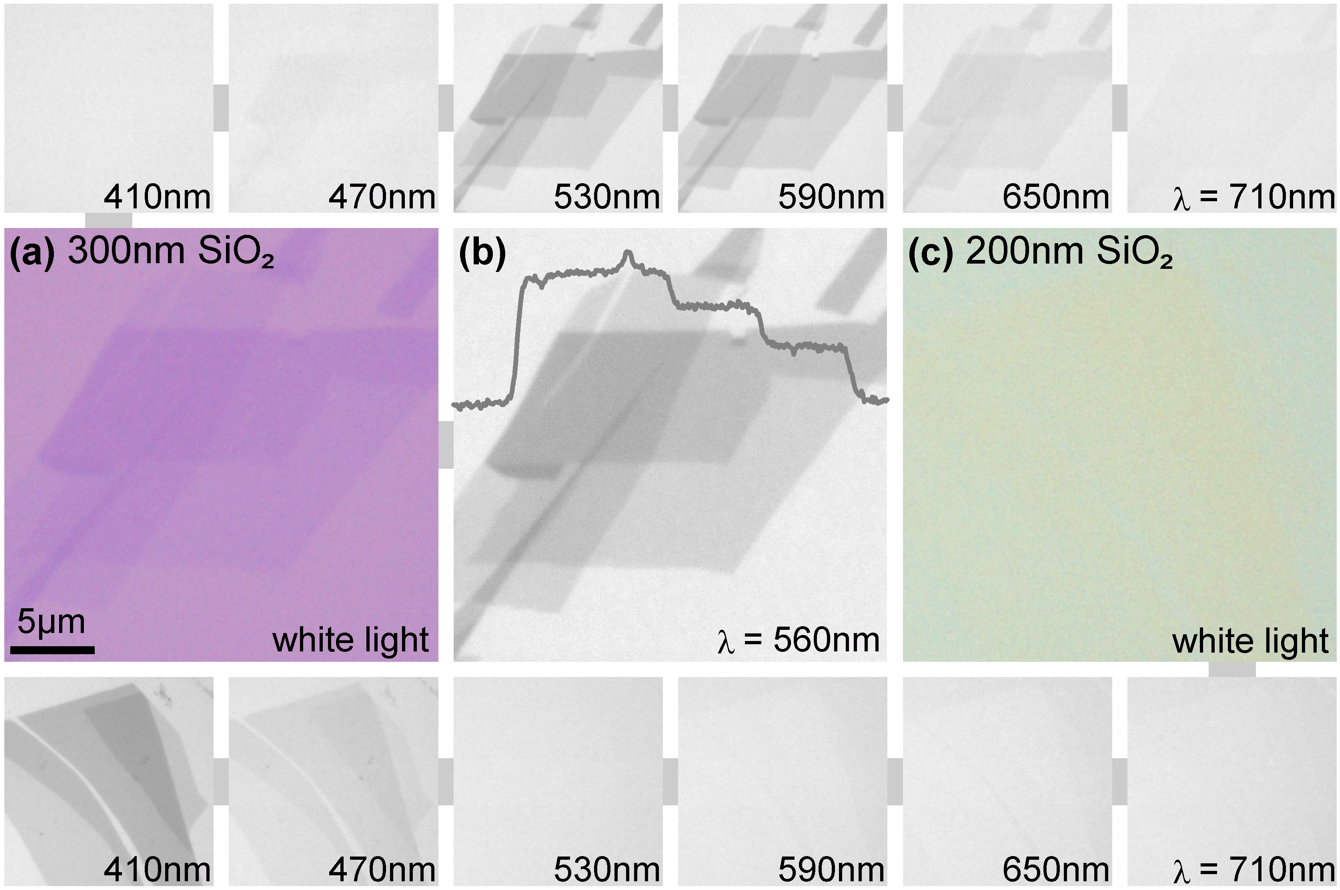}\end{center}

\caption{\label{fig1}(Color online) Graphene crystallites on 300~nm SiO$_2$ imaged with white light (a), green light~\cite{monochrome-cheapest} (b) and another graphene sample on 200~nm SiO$_2$ imaged with white light (c). Single-layer graphene is clearly visible on the left image (a), but even 3 layers are indiscernible on the right (c). Image sizes are 25$\times$25$\mu$m. Top and bottom panels show the same flakes as in (a) and (c), respectively, but illuminated through various narrow bandpass filters with a bandwidth of $\simeq$10~nm.  The flakes were chosen to contain areas of different thickness so that one can see changes in graphene's visibility with increasing numbers of layers. The trace in (b) shows step-like changes in the contrast for 1, 2 and 3 layers (trace averaged over 10-pixel lines). This proves that the contrast can also be used as a quantitative tool for defining the number of graphene layers on a given substrate.}

\end{figure}

In this letter, we discuss the origin of this optical contrast and show that it appears due not only to an increased optical path but also to the notable opacity of graphene. By using a model based on the Fresnel law, we have investigated the dependence of the contrast on SiO$_2$ thickness and light wavelength, $\lambda$, and our experiments show excellent agreement with the theory. This understanding has allowed us to maximize the contrast and, by using narrow-band filters, to find graphene crystallites for practically any thickness of SiO$_2$ and also on other thin films such as Si$_3$N$_4$ and PMMA.

Figure~\ref{fig1} illustrates our main findings. It shows graphene viewed in a microscope (Nikon Eclipse LV100D with a 100$\times$, 0.9 numerical aperture, NA, objective) under normal, white-light illumination on top of a Si wafer with the standard 300nm thickness of SiO$_2$ (Fig.~\ref{fig1}a). For comparison, Fig.~\ref{fig1}c shows a similar sample but on top of 200~nm SiO$_2$, where graphene is completely invisible. In our experience, only flakes thicker than 10 layers could be found in white light on top of 200~nm SiO$_2$. Note that the 10-layer thickness also marks the commonly accepted transition from graphene to bulk graphite~\cite{nmat}. Top and bottom panels in Fig.~\ref{fig1} show the same samples but illuminated through various narrow-band filters. Both flakes are now clearly visible. For 300~nm SiO$_2$, the main contrast appears in green (see Fig.~\ref{fig1}b), and the flake is undetectable in blue light. In comparison, the use of a blue filter makes graphene visible even on top of 200~nm SiO$_2$ (see lower panels).

\begin{figure}[b!]
\begin{center}\includegraphics*[]{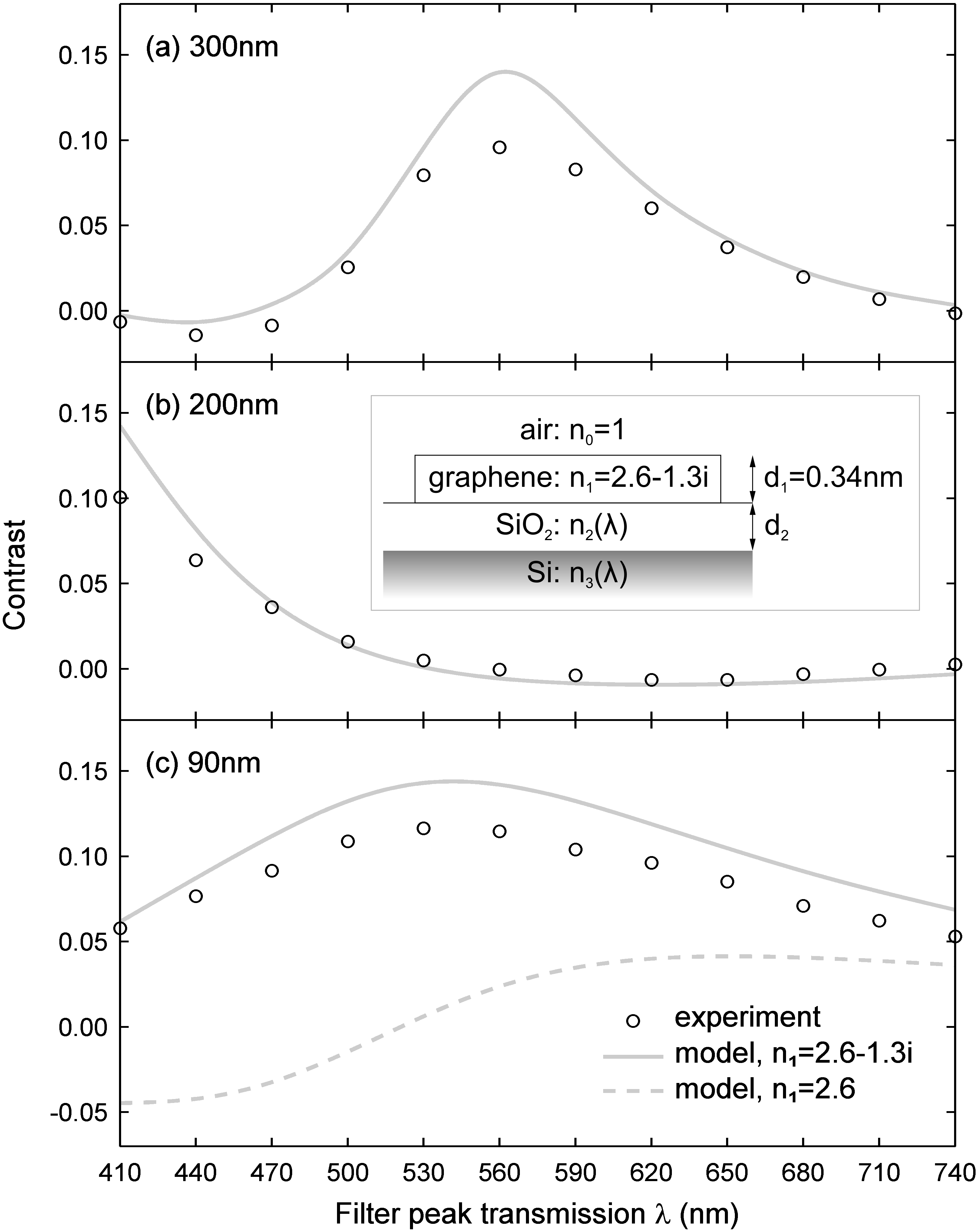}\end{center}
\caption{\label{fig2} Contrast as a function of wavelength for three different thicknesses of SiO$_2$. Circles are the experimental data; curves the calculations. Inset: the geometry used in our analysis.}
\end{figure}

To explain the observed contrast, we consider the case of normal light incidence from air (refractive index, $n_0=1$) onto a tri-layer structure consisting of graphene, SiO$_2$ and Si (see inset of Fig.~\ref{fig2}). The Si layer is assumed to be semi-infinite and characterized by a complex refractive index $n_3(\lambda)$ that, importantly, is dependent on $\lambda$, (for example, $n_3(\lambda = 400 {\rm nm}) \approx 5.6 - 0.4i$)~\cite{palik}. The SiO$_2$ layer is described by thickness $d_2$ and another $\lambda$-dependent refractive index $n_2(\lambda)$ but with a real part only~\cite{palik} ($n_2(400 {\rm nm}) \approx 1.47$). We note that these $n_2(\lambda)$ and $n_3(\lambda)$ accurately describe the whole range of interference colors for oxidized Si wafers~\cite{color-charts}. Single-layer graphene is assumed to have a thickness $d_1$ equal to the extension of the $\pi$ orbitals out of plane~\cite{pauling} ($d_1=0.34$ nm) and a complex refractive index $n_1(\lambda)$. While $n_1(\lambda)$ can be used in our calculations as a fitting parameter, we avoided this uncertainty after we found that our results were well described by the refractive index of bulk graphite $n_1(\lambda) \approx 2.6-1.3i$, which is independent of $\lambda$~\cite{palik, graphite}. This can be attributed to the fact that the optical response of graphite with the electric field parallel to graphene planes is dominated by the in-plane electromagnetic response.

Using the described geometry, it is straightforward to show that the reflected light intensity can be written as~\cite{anders}:
\begin{eqnarray}
I(n_1) &=&
\left| \left(r_1 e^{i (\Phi_1+\Phi_2)} + r_2 e^{-i (\Phi_1-\Phi_2)} \right. \right.
\nonumber
\\
&+& \left. r_3 e^{-i(\Phi_1+\Phi_2)} + r_1 r_2 r_3 e^{i (\Phi_1-\Phi_2)}\right)
\nonumber
\\
&\times & \left(e^{i (\Phi_1+\Phi_2)} + r_1 r_2 e^{-i (\Phi_1-\Phi_2)} \right.
\nonumber
\\
&+& \left. \left.
r_1 r_3  e^{-i(\Phi_1+\Phi_2)} + r_2 r_3 e^{i (\Phi_1-\Phi_2)} \right)^{-1} \right|^2 \, ,
\label{i1}
\end{eqnarray}
\noindent where

\begin{eqnarray}
r_1 = \frac{n_0 - n_1}{n_0+n_1},~r_2 = \frac{n_1 - n_2}{n_1+n_2},~r_3 = \frac{n_2 - n_3}{n_2+n_3}
\end{eqnarray}

\noindent are the relative indices of refraction. $\Phi_1 = 2 \pi n_1 d_1/\lambda$ and $\Phi_2 = 2 \pi n_2 d_2/\lambda$ are the phase shifts due to changes in the optical path. The contrast $C$ is defined as the relative intensity of reflected light in the presence ($n_1\neq 1$) and absence ($n_1=n_0=1$) of graphene:

\begin{eqnarray}
C = \frac{I(n_1=1)-I(n_1)}{I(n_1=1)} \, . \label{contrast}
\label{eq:contrast}
\end{eqnarray}

For quantitative analysis, Fig.~\ref{fig2} compares the contrast observed experimentally with the one calculated by using eq.~(\ref{eq:contrast}). The experimental data were obtained for single-layer graphene on top of SiO$_2$/Si wafers with 3 different SiO$_2$ thicknesses by using 12 different narrow-band filters. One can see excellent agreement between the experiment and theory. The contrast reaches up to $\simeq 12\%$, and the peaks in graphene's visibility are accurately reproduced by our model~\cite{thickness}. Note, however, that the theory slightly but systematically overestimates the contrast. This can be attributed to deviations from normal light incidence (because of high NA) and an extinction coefficient of graphene, $k_1=-\text{Im}(n_1)$, that may differ from that of graphite. $k_1$ affects the contrast both by absoption and by changing the phase of light at the interfaces, promoting destructive interference. To emphasize the important role played by this coefficient, the dashed line in Fig. 2c shows the same calculations but with $k_1=0$. The latter curve does not bare even a qualitative similarity to the experiment, which proves the importance of opacity for the visibility of graphene.

\begin{figure}[h!]
\begin{center}\includegraphics*[]{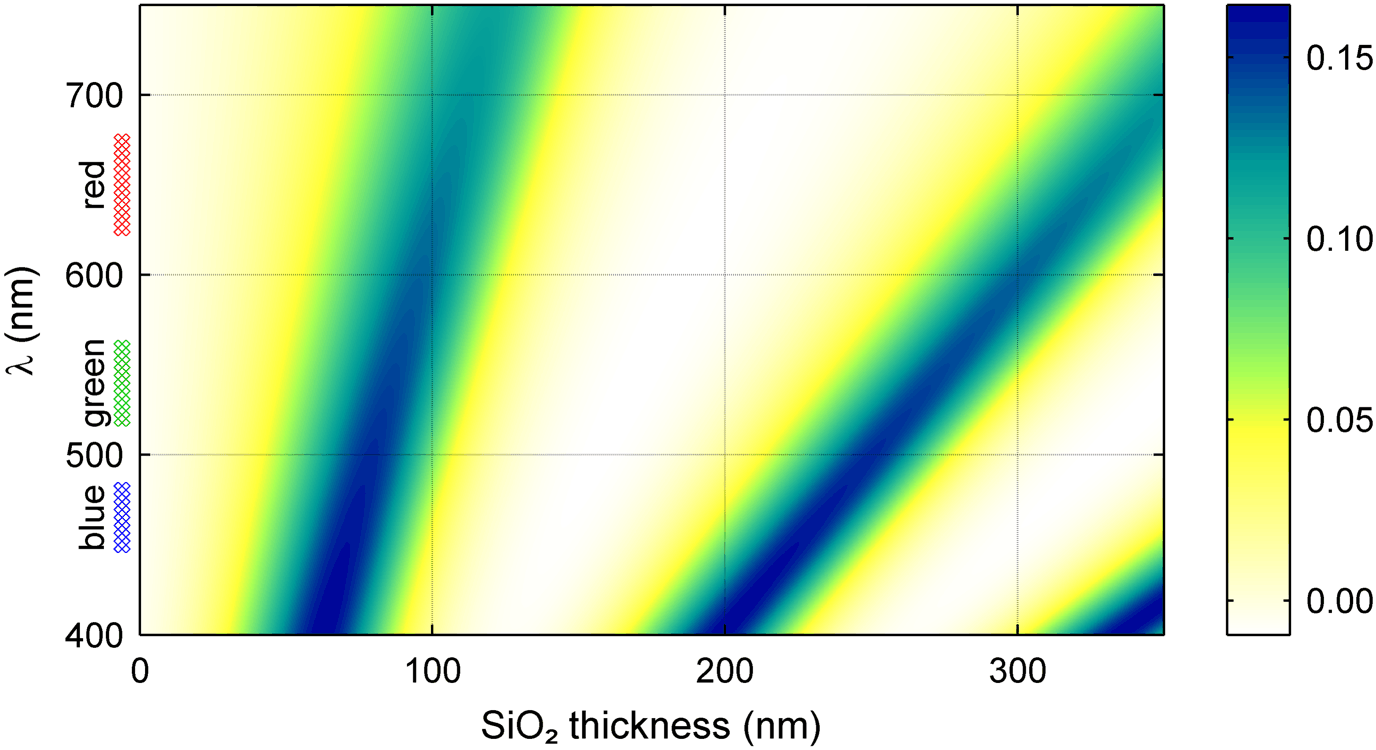}\end{center}
\caption{\label{fig3} (Color online) Color plot of the contrast as a function of wavelength and SiO$_2$ thickness according to eq. (\ref{contrast}). The color scale on the right shows the expected contrast.}
\end{figure}

To provide a guide for the search of graphene on top of SiO$_2$/Si wafers, Fig.~\ref{fig3} shows a color plot for the expected contrast as a function of SiO$_2$ thickness and wavelength. This plot can be used to select filters most appropriate for a given thickness of SiO$_2$. It is clear that by using filters, graphene can be visualized on top of SiO$_2$ of practically any thickness, except for $\approx$150~nm and below 30~nm. Note, however, that the use of green light is most comfortable for eyes that, in our experience, become rapidly tired with the use of high-intensity red or blue illumination. This makes SiO$_2$ thicknesses of approximately 90~nm and 280~nm most appropriate with the use of green filters as well as without any filters, in white light. In fact, the lower thickness of $\simeq$90~nm provides a better choice for graphene's detection (see Fig. 2), and we suggest it as a substitute for the present benchmark thickness of $\simeq$300~nm.

Finally, we note that the changes in the light intensity due to graphene are relatively minor, and this allows the observed contrast to be used for measuring the number of graphene layers (theoretically, multilayer graphene can be modeled by the corresponding number of planes separated by $d_1$). The trace in Fig.~\ref{fig1}a shows how the contrast changes with the number of layers, and the clear quantized plateaus show that we have regions of single, double and triple layer graphene. Furthermore, by extending the same approach to other insulators, we were able to find graphene on 50~nm Si$_3$N$_4$ using blue light and on 90~nm PMMA using white light.

In summary, we have investigated the problem of visibility of graphene on top of SiO$_2$/Si wafers. By using the Fresnel theory, we have demonstrated that contrast can be maximized for any SiO$_2$ thickness by using appropriate filters. Our work establishes a quantitative framework for detecting single and multiple layers of graphene and other 2-dimensional atomic crystals~\cite{pnas} on top of various substrates.

We thank I. Martin for illuminating discussions and C. Luke from Nikon UK for the loan of the monochrome camera~\cite{monochrome-cheapest}. The Manchester work was supported by EPSRC (UK), and A.~H.~C.~N by NSF grant DMR-0343790. After our paper was submitted, four preprints~\cite{same-topic1,same-topic2,same-topic3,same-topic4} discussing the same topic appeared on the cond-mat arXiv.


\begin{thebibliography}{43}
\bibitem{monochrome-cheapest}
Filtered light images are taken with a Nikon DS-2MBWc monochrome camera. White light images are taken with a Nikon DS-2Mv color camera.
Cheaper cameras are more likely to do extensive post-processing of images in firmware or software that could enhance contrast.
\bibitem{graphite}
In Ref.~\cite{palik}, the refractive index of bulk graphite is within 5$\%$ of 2.6-1.3i between 300~nm and 590~nm. At 630~nm, the extinction coefficient jumps to 1.73, but this coincides with a change of reference in the handbook, which we have chosen to ignore in our model.
\bibitem{thickness}
The experimental contrast was found by computer analysis of the images obtained using a monochrome camera~\cite{monochrome-cheapest}. The thickness of SiO$_2$ usually differs by up to 5\% from nominal values provided by suppliers and, accordingly, in our theoretical calculations in Fig.~\ref{fig2}, the following values for $d_2$ were used to acheive the best fit: (a) 290~nm, (b) 190~nm and (c) 88~nm.
\end{thebibliography}
\end{document}